\documentclass[9pt,twocolumn,twoside]{pnas-new}

\templatetype{pnasresearcharticle} 

\usepackage{amsmath}
\usepackage{graphicx}
\usepackage{bm}
\usepackage{mathrsfs}
\usepackage{esint}
\usepackage{enumitem}
\usepackage{multirow}
\usepackage[normalem]{ulem}
\usepackage{xcolor}
\usepackage{hyperref}
\hypersetup{
    colorlinks=true,                          
    linkcolor=red, 
    citecolor=red, 
    urlcolor=red  }
\usepackage{placeins} 
\renewcommand\b[1]{{\bf  #1}}
\renewcommand\vec[1]{\boldsymbol{#1}}

\newcommand\del{\nabla}
\newcommand\e{\epsilon}

\newcommand\dd{\mathrm{d}}

\renewcommand\ss[1]{{\color{red} #1}}

\title{Optimal transport and control of active drops}

\author[a,1]{Suraj Shankar}
\author[b,1]{Vidya Raju}
\author[a,b,c,2]{L.~Mahadevan}

\affil[a]{Department of Physics, Harvard University, Cambridge, MA 02138, USA}
\affil[b]{Paulson School of Engineering and Applied Sciences, Harvard University, Cambridge, MA 02138, USA}
\affil[c]{Department of Organismic and Evolutionary Biology, Harvard University, Cambridge, MA 02138, USA}

\leadauthor{Shankar} 

\significancestatement{Transportation in its broadest sense is an important task in a variety of fields including engineering, physics, biology and economics, and a great deal is known about optimal and efficient strategies to move matter, energy and information around. But can we craft similar optimal protocols to transport autonomously moving (active) matter such as self-propelled active drops or migrating cells? We develop an optimal control framework to transport such active fluid drops with the least amount of energy dissipated, by manipulating the spatio-temporal profile of its internal active stresses. By combining numerical solutions and analytical insight, we uncover simple principles and characteristic trade-offs that govern the optimal policies, suggesting general strategies for optimal transportation in a wide variety of synthetic and biological active systems.}

\authorcontributions{}
\authordeclaration{The authors declare no competing interests.}
\equalauthors{\textsuperscript{1}S.~S.~and V.~R.~contributed equally to this work.}
\correspondingauthor{\textsuperscript{2}Correspondence and requests for materials
should be addressed to L.~M.~Email: lmahadev@g.harvard.edu~.}


\begin{abstract}
Understanding the complex patterns in space-time exhibited by active systems has been the subject of much interest in recent times. Complementing this forward problem is the inverse problem of controlling active matter. Here we use optimal control theory to pose the problem of transporting a slender drop of an active fluid and determine the dynamical profile of the active stresses to move it with minimal viscous dissipation. By parametrizing the position and size of the drop using a low-order description based on lubrication theory, we uncover a natural ``gather-move-spread'' strategy that leads to an optimal bound on the maximum achievable displacement of the drop relative to its size. In the continuum setting, the competition between passive surface tension, and active controls generates richer behaviour with futile oscillations and complex drop morphologies that trade internal dissipation against the transport cost to select optimal strategies. Our work combines active hydrodynamics and optimal control in a tractable and interpretable framework, and begins to pave the way for the spatiotemporal manipulation of active matter.
\end{abstract}

\dates{This manuscript was compiled on \today}
\doi{\url{www.pnas.org/cgi/doi/10.1073/pnas.XXXXXXXXXX}}

\begin{document}

\maketitle
\thispagestyle{firststyle}
\ifthenelse{\boolean{shortarticle}}{\ifthenelse{\boolean{singlecolumn}}{\abscontentformatted}{\abscontent}}{}

\dropcap{I}n recent years, active fluids composed of internally driven units have emerged as a powerful platform to manipulate and morph matter far from equilibrium \cite{needleman2017active,zhang2021autonomous,shankar2020topological}. Such fluids have been assembled from a variety of biological and synthetic constituents, including self-propelled colloids, driven biofilaments and living cells \cite{sanchez2012spontaneous,bricard2013emergence,zhou2014living,duclos2018spontaneous}. These systems often exhibit complex spatiotemporal dynamics and pattern formation that has been the focus of intense research efforts in the past two decades or so.

While a great deal is now understood about the emergent collective dynamics in active fluids \cite{marchetti2013hydrodynamics}, much less is known about how we can control or harness such collective phenomena to achieve functional goals. Recent experimental advances in microfabrication and optogenetic techniques have allowed the development of novel bacterial and synthetic reconstituted systems to begin addressing this question in different contexts, such as active engines for efficient work extraction \cite{vizsnyiczai2017light,krishnamurthy2016micrometre}, the dynamic control of reconfigurable density patterns \cite{arlt2018painting,frangipane2018dynamic}, and the targeted creation and transport of localized structures such as defects \cite{ross2019controlling,zhang2021spatiotemporal}. On a different scale, the collective control of migrating and proliferating cellular monolayers through patterned substrates \cite{turiv2020topology,endresen2019topological} or external fields \cite{cohen2014galvanotactic,zajdel2020scheepdog} also presents new possibilities for the control of active biological matter.

The capacity for spontaneous and autonomous motion in active fluids raises a natural question: what are the optimal strategies to spatially transport active materials? The general problem of optimal mass transport, i.e., finding the easiest or cheapest way to move mass from one place to another, began with early work by Monge \cite{monge1781memoire} and Kantorovich \cite{kantorovich1942translocation}, and has been explored extensively in mathematics, economics and the physical sciences for over two centuries now \cite{villani2008optimal}. While the conventional Monge-Kantorovich formulation of optimal transport relies on a single constraint of global mass conservation and eschews any explicit dynamics, Benamou and Brenier \cite{benamou2000computational} converted the problem to one of trajectory optimization by reformulating its solution in terms of fluid dynamics, albeit of a fictional inviscid fluid obeying pressure-less potential flow. But real fluids, including active fluids, obey complex physical dynamics without immediate counterparts in standard optimal transport formulations, leading us to ask, how can we construct optimal transport policies to move physical materials?

Here we pose this question in the simplest setting of transporting an active drop by dynamically controlling its internal activity in space-time. By incorporating the dynamical constraints of droplet motion using the lubrication approximation, and expressing the cost of transport in terms of the dissipation rate, we ask if we can determine the internal activity to move the drop from one place to another while minimizing the total dissipation, thus bringing it within the framework of optimal control theory.   

\begin{figure*}[t]
	\begin{center}
	\includegraphics[width=\textwidth]{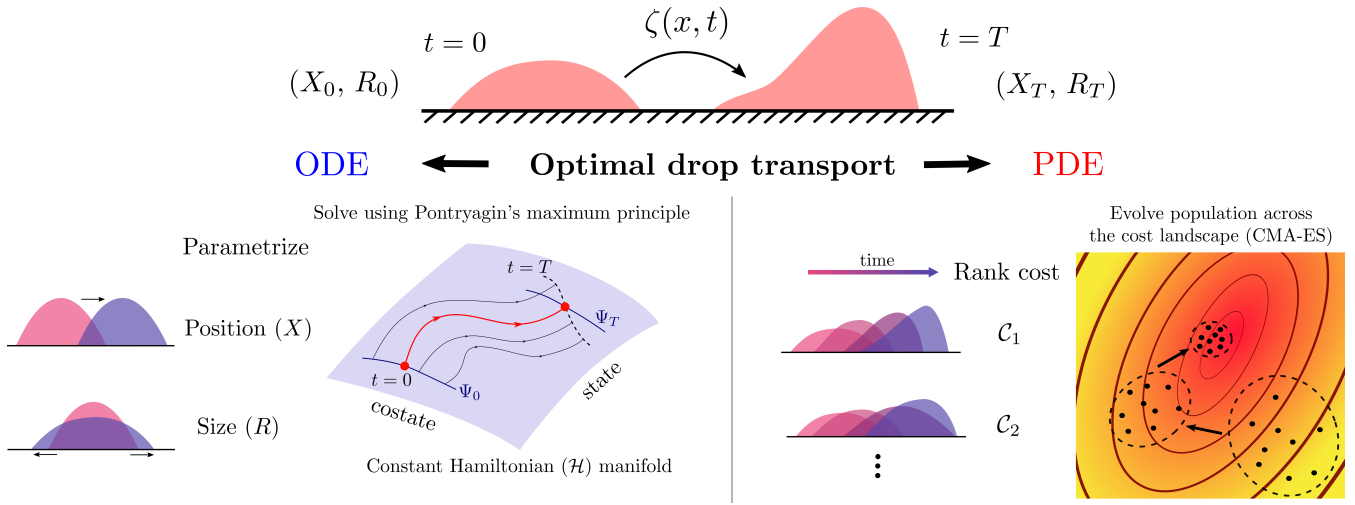}
	\end{center}
	\caption{\textbf{Optimal transport of an active drop.} A schematic illustrating our framework to solve the problem of transporting an active drop by minimizing a specified cost function, such as the mechanical work. The spatiotemporal profile of activity $\zeta(x,t)$ is the control variable and the transport task involves moving the drop from position $X_0$ and size $R_0$ to a final position $X_T$ and size $R_T$ in a finite time $T$. We employ two complementary approaches - (i) finite dimensional optimal control using Pontryagin's maximum principle on an ODE based reduced order model that capture parametrized features of the drop, (ii) constrained numerical optimization of the nonlinear continuum PDE using a gradient-free evolutionary algorithm such as Covariance Matrix Adaptation (CMA-ES, see main text).}
	\label{fig:odepde}
\end{figure*}

By projecting the continuum description of droplet motion onto a finite dimensional slow manifold, we derive a reduced system of ordinary differential equations for the position and size of the droplet. Interestingly, for a range of parameters,  an intuitive ``gather-move-spread'' style strategy emerges naturally as an optimum within our framework. Numerical simulations of the continuum equations using an evolutionary algorithm to determine the optimal activity profile confirms the qualitative nature of the results obtained from the reduced order model. Together these solutions provide a comprehensive yet interpretable framework (Fig.~\ref{fig:odepde}) to understand the optimal solutions obtained, and highlights the resulting trade-offs between cost, transport precision and efficiency that can be tuned by the interplay of passive and active stresses in the system. 

\section*{Mathematical model for optimal droplet transport}
\subsection*{Lubrication dynamics of an active drop}
\label{sec:drop}
\begin{figure}[]
	\begin{center}
	\includegraphics[width=0.5\textwidth]{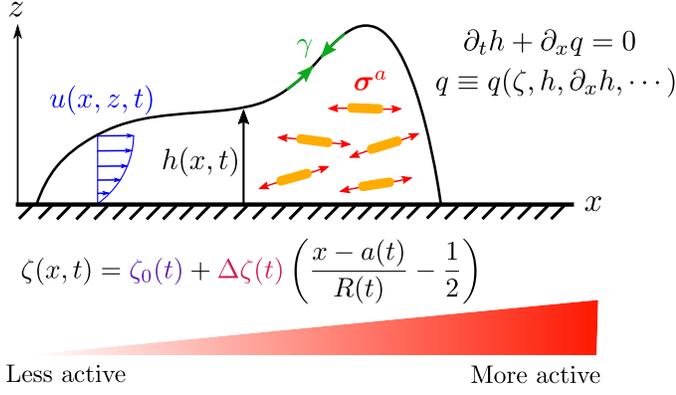}
	\end{center}
	\caption{\textbf{Model of an active drop moving on a substrate.} The horizontal flow velocity $u(x,z,t)$ driven by active internal stresses ($\vec{\sigma}^a$) and surface tension ($\gamma$) adopts a Poiseuille-like profile in the drop interior, characteristic of lubrication theory. The drop height $h(x,t)$ obeys the continuity equation (Eq.~\ref{eq:cont}) with a flux $q$ that encodes the constitutive relation (Eq.~\ref{eq:q}) for how activity drives fluid flow. The spatial profile of activity ($\zeta(x,t)$) is a simple linear ramp with a constant offset, allowing for both drop translation and size change (Eq.~\ref{eq:zeta}).}
	\label{fig:drop}
\end{figure}

We describe the dynamics of a slender drop of an active suspension on a solid surface in the asymptotic limit exemplified by viscous lubrication theory \cite{joanny_ramaswamy_2012,loisy2019tractionless,*loisy2020many}. For simplicity, we consider a two-dimensional drop (2D) moving in the $x$-direction (see Fig.~\ref{fig:drop}) and neglect gravity by assuming the drop size to be smaller than the capillary length. Fluid incompressibility requires that $\vec{\del}\cdot\b{u}=\partial_xu+\partial_zv=0$, where $\b{u}=(u(x,z,t),v(x,z,t))$ is the local flow velocity. Upon depth integrating the incompressibility equation, and noting the free surface boundary condition $v|_{z=h}=\partial_th+u\partial_xh|_{z=h}$, where $h(x,t)$ is the height profile of the drop, we obtain the conservation law
\begin{equation}
	\partial_th+\partial_xq=0\;.\label{eq:cont}
\end{equation}
Here the mass flux $q=h\langle u\rangle$, with the average horizontal velocity $\langle u\rangle=(1/h)\int_0^h\dd z\;u$.
Assuming that there is no addition or loss of mass of the drop, we can write this as a global condition
\begin{equation}
	\int\dd x\;h(x,t)=1\;,\label{eq:hint}
\end{equation}
to fix our units of length.

In the low-Reynolds number regime appropriate for small viscous drops, we operate in the Stokesian limit, wherein force balance implies $\vec{\del}\cdot\vec{\sigma}=\b{0}$, where the total stress $\vec{\sigma}=-p\b{I}+\eta[\vec{\del}\b{u}+(\vec{\del}\b{u})^T]+\vec{\sigma}^a$ is the sum of the pressure $p$, a viscous stress (in a liquid with shear viscosity $\eta$) and an active stress. We assume that the active stress $\vec{\sigma}^a= \zeta h(\hat{\b{n}}\hat{\b{n}} - \b{I}/2) $ 
\cite{simha2002hydrodynamic,marchetti2013hydrodynamics} is proportional to the local density $\sim h$, while depending on the local orientation $\hat{\b{n}}$ of anisotropic active agents\footnote{An isotropic active pressure that is constant across the thickness of the drop does not change anything as it can be absorbed into $p$.}. The activity $\zeta(x,t)$ depends on space and time and originates from the density of force dipoles exerted by elongated active units, which can be of either sign, with $\zeta>0$ for contractile systems and $\zeta<0$ for extensile systems. This form of the active stress is applicable to drops of coherently swimming dense bacterial suspensions or ordered collections of motor protein driven cytoskeletal filaments present in synthetic drops or living cells \cite{joanny_ramaswamy_2012,loisy2019tractionless}. For simplicity, we shall assume strong ordering along the horizontal direction and neglect any rapid orientational relaxation to set $\hat{\b{n}}\simeq\hat{\b{x}}$ to lowest order in gradients ($|\partial_xh|\ll 1$). The active forcing nonetheless survives in this limit, as the active stress directly depends on the local density of the drop ($\sim h$), unlike previous models \cite{joanny_ramaswamy_2012,loisy2019tractionless} that rely on splay-bend deformations of orientational order.

In the lubrication limit corresponding to a slender drop, $|\partial_xh|^2 \ll 1$ so that we can neglect longitudinal flow gradients as $|\partial_x^2u|\ll|\partial_z^2u|$ and $|\partial_xv|\ll|\partial_zu|$. Then the stress in the fluid is dominated by the pressure, which is determined by the capillary boundary condition on the drop surface $\sigma_{zz}|_{z=h}=\gamma\partial_x^2h$, where $\gamma$ is the interfacial tension, and yields  $p=-(\zeta/2)h-\gamma\partial_x^2h$.   Using this in the horizontal force balance equation, along with the no-slip ($u|_{z=0}=0$) and free surface ($\partial_zu|_{z=h}=0$) boundary conditions, yields the horizontal velocity profile $u=z(2h-z)\partial_x\sigma/2\eta$ where $\sigma=\zeta h+\gamma\partial_x^2h$ is the effective uniaxial stress. Averaging the velocity through the thickness of the drop shows that the net flux is
\begin{equation}
	q=h\langle u\rangle=\dfrac{h^3}{3\eta}\partial_x(\zeta h+\gamma\partial_x^2h)\;.\label{eq:q}
\end{equation}
Eqs.~\ref{eq:cont} and~\ref{eq:q} complemented by boundary and initial conditions on the height of the film and its derivatives completely describe the macroscopic dynamics of an active drop on a substrate, once the activity field $\zeta(x,t)$ is specified. For finite drops, in the neighborhood of the contact line the boundary conditions associated with partial slip, pre-wetting films and/or finite contact angles have to be accounted for \cite{de1985wetting,de2013capillarity} (see SI). In our formulation,  the passive surface tension $\gamma$ serves to regulate the drop curvature, while the (controllable) activity $\zeta(x,t)$ enters as an unknown spatiotemporally varying nonlinear diffusivity, but is analogous to gravity \cite{de1985wetting}.

\subsection*{Optimal transport}
\label{sec:opt}
The optimal transport of an active drop requires finding an actuation protocol for the activity profile $\zeta(x,t)$ that moves the drop at a minimal cost. We choose a physically motivated cost $\mathcal{C}=\mathcal{W}+\mathcal{T}$ which includes two terms, an integrated bulk cost that tracks the total mechanical work ($\mathcal{W}$) done by the active and passive forces, and a terminal cost ($\mathcal{T}$) that penalizes any discrepancy between the final and desired state of the drop. Within the lubrication approximation, the viscous dissipation in the drop $\sim \eta(\partial_zu)^2$ is dominated by shear, so that the total mechanical work is given by
\begin{equation}
	\mathcal{W}=\int_0^T\dd t\int\dd x\;\dfrac{h^3}{3\eta}(\partial_x\sigma)^2\;.\label{eq:W}
\end{equation}
As expected, the total amount of energy expended and lost via dissipation by the system is always non-negative ($\mathcal{W}\geq 0$). We note that this is equivalent to stating that the effective mechanical energy in the drop $E = (1/2)\int\dd x[\gamma (\partial_xh)^2-\zeta h^2]$ satisfies the condition $\dd E/\dd t = -\int\dd x (h^3/3\eta)(\partial_x\sigma)^2<0$ if $\partial_t\zeta=0$ and boundary fluxes are absent, i.e., the system is dissipative. The work done by the active stress alone ($\mathcal{W}_a=\int_0^T\dd t\int\dd x\;h\langle u\rangle\partial_x(\zeta h)$) on the other hand is not guaranteed to be a well-behaved cost function as $\mathcal{W}_a$ can be of either sign in general (though $\mathcal{W}_a=\mathcal{W}\geq 0$ when $\gamma=0$), reflecting the possibility of both energy consumption and extraction from the nonequilibrium system \cite{pietzonka2019autonomous}.
Here, we have not included the energy cost required to \emph{maintain} the active machinery \cite{recho2014optimality}; in the simplest setting this is proportional to the total amount of the active fluid, which in our case is a constant. 

For simplicity, we have assumed that the total time duration $T$ is fixed, though, other strategies such as minimal time control are possible. However, we do account for a terminal cost to minimally captures the intent of the task, which is to translate the drop by a fixed distance and control its final spread as well. We incorporate this in a simple quadratic term
\begin{equation}
	\mathcal{T}=\mu_X\left(\dfrac{X(T)-X_T}{X_T}\right)^2+\mu_R\left(\dfrac{R(T)-R_T}{R_T}\right)^2\;,\label{eq:T}
\end{equation}
where $X_T$ and $R_T$ are the desired values for the drop center of mass and the drop size at the end of the transport. The corresponding penalties are $\mu_X$ and $\mu_R$ for the terminal position and size terms. As mentioned previously, the drop has compact support and finite size which is denoted by $R(t)$ and its position is given by the center of mass, namely
\begin{equation}
	X(t)=\int\dd x\;xh(x,t)\;,
\end{equation}
both of which can be evaluated at the final time $t=T$ to compute $\mathcal{T}$ (Eq.~\ref{eq:T}). We will always set the initial position of the drop to be at the origin, $X(0)=0$, without loss of generality.

It is worthwhile to pause here to compare our formulation of optimal droplet transport with the classical Monge-Kantorovich formulation of optimal transport \cite{monge1781memoire,kantorovich1942translocation,benamou2000computational,villani2008optimal}. Unlike the conventional formulation, where the sole constraint is global mass conservation (Eq.~\ref{eq:hint}), here we constrain the dynamics to account for both \emph{local} mass and momentum conservation. The latter is a direct consequence of the physics of fluid motion that dictates how the material responds to local actuation of active stresses, as specified by Eqs.~\ref{eq:cont} and~\ref{eq:q}. As a result, our transport plan does not simply rely on a registration solution of a static Monge-Amp{\'e}re equation \cite{villani2008optimal}.
A further salient feature worth emphasizing is the parabolic nature of our dynamics where the control (activity) enters as a nonlinear diffusivity by virtue of the geometric reduction intrinsic to drops and thin films. This is distinct from the hyperbolic setting present in Benamou-Brennier style formulations \cite{benamou2000computational} where the velocity field is the control variable. 


\section*{Protocol for optimal droplet transport}
With this minimal physical framework in hand, how can we compute the optimal transport policies? We choose two different routes of solving the problem (Fig.~\ref{fig:odepde}). First, we project our nonlinear partial differential equation (PDE) for the drop dynamics (Eqs.~\ref{eq:cont},~\ref{eq:q}) onto a finite number of low order modes that correlate with the location, size and shape of the droplet. We propose a strategy that minimizes or eliminates drift terms in the reduced description by considering the active controlled dynamics within a low-dimensional slow manifold that is approximately invariant to the passive (non-controlled) forces. The resulting ordinary differential equations (ODEs) can then be handled using standard optimal control theory \cite{liberzon2011calculus}, which we solve both analytically and numerically. We then compare this reduced order description to the full PDE model, for which we analyze the optimal control problem numerically. However, unlike conventional forward problems that are solved with initial values in time, the controlled dynamics requires the solution to a much harder boundary-value-problem in time. One way to solve this problem is to stochastically search for different initializations using a large number of forward runs to find an appropriate solution satisfying the required boundary condition. We implement this using a gradient-free covariance matrix adaptation evolution strategy (CMA-ES) \cite{hansen2006cma}, as explained later and in the SI.  

\subsection*{ODE control}
\label{sec:ode}
\begin{figure*}[t]
	\begin{center}
		\includegraphics[width=\textwidth]{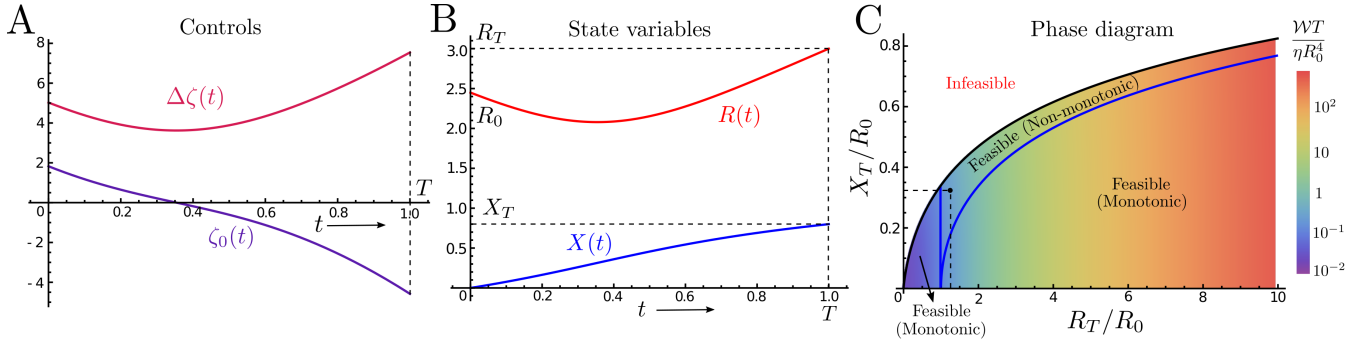}
	\end{center}
	\caption{\textbf{ODE optimal control.} (A-B) Sample trajectories for the globally optimal mean ($\zeta_0$) and gradient ($\Delta\zeta$) activity are shown in A, and the associated controlled dynamics for the drop position ($X$) and size ($R$) are shown in B. The parameters chosen are $X_T=0.8$, $R_0=\sqrt{6}$, $R_T=3$, $T=1$ and $\eta=0.1$. Note that as $\eta$ sets a time scale, only the ratio $T/\eta$ is important. For these parameter values, we see that the size change is non-monotonic, which is reflected in the sign change in the mean activity $\zeta_0(t)$. The initial contractile activity ($\zeta_0>0$) causes the drop to shrink and consequently accelerate its translation, and later the activity switches over to become extensile ($\zeta_0<0$) to allow the drop to reach its larger final size $R_T$. (C) The phase diagram here represents the feasibility region for optimal transport of a parabolic active drop, as a function of the non-dimensionalized drop displacement ($X_T/R_0$) and the its size disparity ($R_T/R_0$). The black curve is the maximum achievable displacement $X_T$ for a given relative change in size ($R_T/R_0$), beyond which no smooth optimal controls exist. Below the black curve is the feasible region, with the shaded colour representing the total work done (nondimensionalized as $\mathcal{W}T/(\eta R_0^4)$) by the globally optimal policy, with the cost increasing from blue to red. The blue curves in the shaded region demarcate the parameter space where the global optimizer has a monotonic or non-monotonic size dependence as a function of time. Non-monotonic changes in size are favoured in the region bordered by the blue and black curves, and only occur when $R_T>R_0$. For $R_T<R_0$, the optimal policies have a monotonic size dependence. The black dot corresponds to the solution shown in (A-B).}
	\label{fig:sym}
\end{figure*}

For simplicity, we consider a minimal setting where the spatial variation of the activity is fixed, but its time variation is arbitrary. As we are primarily interested in controlling the position and size of the drop, the simplest spatial variation of activity that can accomplish both is a linear profile,
\begin{equation}
	\zeta(x,t)=\zeta_0(t)+\Delta\zeta(t)\left(\dfrac{x-a(t)}{R(t)}-\dfrac{1}{2}\right)\;,\label{eq:zeta}
\end{equation}
where $\zeta_0(t)$ is a time-varying mean activity  and $\Delta\zeta(t)$ is a time-varying gradient in activity  (Fig.~\ref{fig:drop}). We choose this specific form which depends explicitly on the drop size $R(t)$ and the position of the left edge of the drop $a(t)$, so that the (spatial) average activity in the drop is $(1/R(t))\int_{h>0}\dd x\;\zeta(x,t)=\zeta_0(t)$. We note that while $\zeta_0$ essentially controls the size of the drop, with $\zeta_0>0$ leading to contraction and $\zeta_0<0$ leading to expansion, the linear gradient in activity $\Delta\zeta$ controls the drop translation, the direction of which is determined by the sign of $\Delta\zeta$. In contrast with recent works that have explored bulk contractility driven crawling of cells \cite{recho2013contraction,loisy2020many} and shown how it optimizes the mechanical efficiency of steady motion \cite{recho2014optimality}, here we address the unsteady dynamics and its control problem.

The relative importance of the active drive versus surface tension is quantified by a dimensionless \emph{active} capillary number
\begin{equation}
	{\rm Ca}_\zeta=\left\langle\dfrac{|\Delta\zeta(t)|R(t)^2}{\gamma}\right\rangle_T\;,\label{eq:Ca}
\end{equation}
where the time average $\langle A\rangle_T=(1/T)\int_0^T\dd t\;A$. As will be clear later, Eq.~\ref{eq:Ca} is akin to the conventional definition of a capillary number \cite{de2013capillarity}, only now with the velocity scale set by the activity gradient ($\Delta\zeta$) which is essential to drive drop motion\footnote{Similarly, we can define an \emph{active} Bond number ${\rm Bo}_\zeta=\langle|\zeta_0(t)|R(t)^2/\gamma\rangle_T$ to characterize the relative importance of the average activity (associated with size change) compared to surface tension. Since $\rm{Ca}_\zeta$ is more directly relevant for transport and in most of our results, for instance in Fig.~\ref{fig:pde}, both $\rm{Bo}_\zeta$ and $\rm{Ca}_\zeta$ are empirically correlated (not shown), we do not consider the active Bond number any further.}. For large Ca$_{\zeta}\gg 1$, active forcing dominates surface tension and we can safely neglect boundary effects, while for Ca$_\zeta\sim 1-\mathcal{O}(10)$, active and passive forces compete and the equations have to be generalized to include the dynamics of the contact lines, as detailed in the SI.

We project the nonlinear PDE (Eqs.~\ref{eq:cont} and~\ref{eq:q}) onto a truncated set of modes that we choose so that the resulting ODE system is as drift-free as possible, i.e., the system lacks dynamics in the absence of the controls (here, activity). This is achieved by noting that the flux due to capillary forces vanishes when $\partial_x^3h(x,t)=0$ and the drop adopts a parabolic profile. This permits a simple parametrization of the drop profile via two modes of deformation - a translation in the center of mass $X(t)$ and a change in the size $R(t)$, which along with Eq.~\ref{eq:hint} gives
\begin{equation}
	h(x,t)=\dfrac{6}{R(t)^3}\left[\dfrac{R(t)^2}{4}-\left(x-X(t)\right)^2\right]\;.\label{eq:hparam}
\end{equation}
Note that $h(x,t)=0$ at the two ends of the drop, $x=X(t)\pm R(t)/2$, and vanishes outside this region. While translation is a genuine zero mode of capillarity, size change of the drop is only an approximate zero mode that is violated near the boundaries where wetting and contact angle physics becomes important \cite{de2013capillarity}. Nonetheless, by focusing on the bulk dynamics, we obtain a two-dimensional manifold spanned by $X(t)$ and $R(t)$ that remains approximately invariant under the action of capillary forces. 

Employing a Galerkin approximation allows us to compute spatial moments of the flux $q(x,t)$ and project the continuum equations onto this manifold to obtain a pair of nonlinearly coupled ODEs (see SI)
\begin{equation}
	\dot{X}(t)=\dfrac{18\Delta\zeta(t)}{35\eta R(t)^4}\;,\quad \dot{R}(t)=-\dfrac{24\zeta_0(t)}{7\eta R(t)^4}\;.\label{eq:XR}
\end{equation}
As expected, the mean ($\zeta_0$) and gradient ($\Delta\zeta$) components of the active stress independently control the drop size and position respectively. By construction, surface tension $\gamma$ is absent in Eq.~\ref{eq:XR} and the equation is drift-free. The existence of optimal controls is predicated on an important property of the dynamics, namely \emph{controllability} \cite{brockett1976nonlinear,hermann1977nonlinear}, i.e. the existence of a path connecting any two points in the state or configuration space, spanned here by $X$ and $R$ (see SI for a more detailed explanation). As the two controls ($\zeta_0,\Delta\zeta$) enter linearly and independently, and the dynamics has no fixed points for nonvanishing controls, one can easily confirm that Eq.~\ref{eq:XR} is controllable, allowing us to guarantee the ability to steer the system from any state to any other state within the space of drop configurations labeled by $(X,R)$. The absence of any drift (control independent) terms presents a technical advantage as the system permits a global rather than local notion of controllability even when the dynamics is nonlinear (see SI), thereby justifying our mode reduction strategy.

For simplicity, we consider the fixed end point problem where the terminal conditions are imposed strictly ($X(T)=X_T$, $R(T)=R_T$), in which case, the total cost reduces to the net dissipation ($\mathcal{C}=\mathcal{W}$). The drop parametrization (Eq.~\ref{eq:hparam}) allows us to easily compute the dissipation rate to be
\begin{equation}
	\mathcal{L}=\dfrac{1}{\eta R^6}\left[\dfrac{72}{35}\zeta_0^2+\dfrac{54}{77}\Delta\zeta^2\right]\;,\label{eq:L}
\end{equation}
whose time integral gives the cost ($\mathcal{W}=\int_0^T\dd t\;\mathcal{L}$). As expected, we obtain a simple sum of squares in terms of the two active drives, along with a strong size dependence arising from the geometry of the drop. To solve the optimal control problem, we employ Pontryagin's Maximum Principle that provides the necessary conditions for optimality \cite{pontryagin2018mathematical}. Upon introducing the costates (Lagrange multipliers) $p_X(t)$ and $p_R(t)$ to enforce the dynamical constraints in Eq.~\ref{eq:XR}, a necessary condition for optimality is the maximization of the control Hamiltonian
\begin{equation}
	\mathcal{H}=p_X\dfrac{18\Delta\zeta}{35\eta R^4}-p_R\dfrac{24\zeta_0}{7\eta R^4}-\mathcal{L}\;,\label{eq:cH}
\end{equation}
with respect to the controls (see SI). This gives $\zeta^*_0=-5p_RR^2/6$ and $\Delta\zeta^*=11p_XR^2/30$, which when substituted back into Eq.~\ref{eq:cH} gives the conserved Hamiltonian $H=\mathcal{H}(\zeta_0^*,\Delta\zeta^*)$.
The candidate extremals for the optimal control problem satisfy Hamiltonian dynamics in terms of the state variables ($\dot{X}=\partial_{p_X}H$, $\dot{R}=\partial_{p_R}H$) and corresponding costates ($\dot{p}_X=-\partial_XH$, $\dot{p}_R=-\partial_RH$). For the state variables, this gives back Eq.~\ref{eq:XR} driven now by the optimal controls $(\zeta_0^*,\Delta\zeta^*)$; translational invariance enforces that $\partial_X\mathcal{H}=0$, hence $p_X$ is conserved along the optimal trajectory.

These coupled dynamical equations along with the initial and terminal conditions can be solved analytically to obtain the optimal transport protocols (see SI) to displace an active drop by a distance $X_T$ and change its size from $R_0$ to $R_T$ in a finite time interval $T$. A representative solution is plotted in Fig.~\ref{fig:sym}(A-B) for $X_T=0.8$, $R_0=\sqrt{6}$ and $R_T=3$. The first order necessary conditions generally allow for non-uniqueness of the candidate extrema, but we only show the global optimizer in Fig.~\ref{fig:sym}. For the chosen parameters, the optimal protocol leads to a non-monotonic change in drop size (Fig.~\ref{fig:sym}B), first decreasing and later increasing to reach the final size $R_T$. This is reflected in the sign change of the mean active stress $\zeta_0$ (Fig.~\ref{fig:sym}A), which switches from being contractile initially ($\zeta_0>0$) to extensile at later times ($\zeta_0<0$). The drop translates smoothly with a maximal velocity precisely when the drop size is the smallest, even though the drive ($\Delta\zeta$) is minimal at this point (Fig.~\ref{fig:sym}A). Hence, the drop executes a continuous version of an intuitive ``gather-move-spread'' like strategy that naturally emerges as an optimal transport plan in our framework. A simple concentration effect that enhances the active drive in smaller drops underlies this phenomenon by allowing for faster transport at lower activity. Our solution reveals a further striking result - for certain values of the target parameters (set by $X_T$ and $R_T$) with strict terminal constraints, \emph{no} continuous optimal policies for the transport problem exist! This does not contradict the fact that the dynamical system (Eq.~\ref{eq:XR}) is controllable, but rather highlights a subtlety. While controllability guarantees the presence of a trajectory in configuration space that steers the drop to its desired final state, and hence the existence of optimal controls, it does not in general guarantee this transport is either continuous or achievable in finite time.

As shown by the coloured region bounded by the black curve in Fig.~\ref{fig:sym}C, we have a feasible or reachable regime where \emph{smooth} optimal solutions exist and the net dissipation is finite (shown in the heat map, with the cost increasing from blue to red), while in the infeasible region, no smooth solution satisfies the terminal conditions. The blue curve in Fig.~\ref{fig:sym}C further demarcates the parameter regime where the global optimum corresponds to polices with a monotonic change in drop size, where the mean activity ($\zeta_0$) maintains a fixed sign. These appear either for sufficiently small displacements or when the final drop size is smaller than the initial one ($R_T<R_0$). Nonsmooth protocols can be constructed to access points in the infeasible region in Fig.~\ref{fig:sym}C, but they lack a natural parametrization. As a result, the search for an optimal policy in this larger function space is analytically intractable and we focus only on smooth protocols for simplicity. The feasibility curve can alternatively be viewed as solving a maximin problem, where we maximize the minimum dissipation protocol over the translation $X_T$ for a fixed change in drop size ($R_T/R_0$), the solution for which roughly equipartitions the transport cost (black curve in Fig.~\ref{fig:sym}C, see SI for further details).

Extending the ODE analysis to include changes in the drop shape via an asymmetry in the height profile (see SI for details) and solving the resulting equations using CasADi \cite{andersson2019casadi} and homotopy continuation techniques, we obtain optimal solutions that progressively deviate from the analytical solution (Fig.~\ref{fig:sym}) computed for the symmetric drop (Eq.~\ref{eq:hparam}). While small variations in the drop asymmetry result in smooth optimal policies, that are similar to those in Fig.~\ref{fig:sym}A, larger values of drop asymmetry lead to solutions that depend sensitively on the terminal constraints and exhibit sharp jumps in the dynamics of the drop configuration that are not easy to interpret (see SI). This suggests that a simple Galerkin-approximation and truncation beyond two modes (the position and size) requires more care and is susceptible to spurious instablities that do not faithfully capture the full continuum dynamics of the drop (as will be evident in the following section). In light of this, we switch to the full PDE model.

\subsection*{PDE control}
\label{sec:pde}
\begin{figure*}[t]
	\centering{\includegraphics[width=\textwidth]{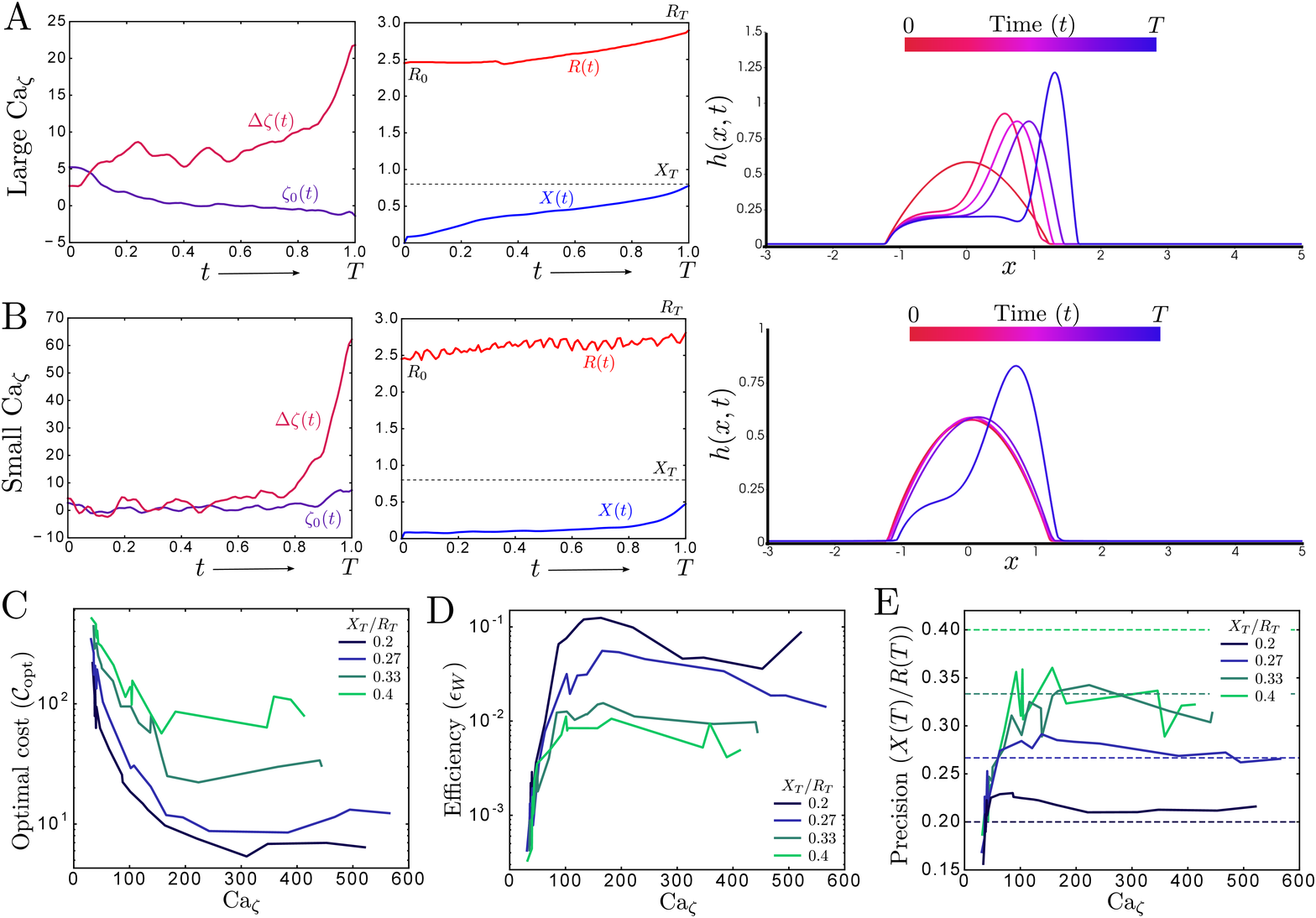}}
	\caption{\textbf{PDE optimal control.} (A) The optimal activity controls ($\zeta_0(t),\,\Delta\zeta(t)$, left) and corresponding trajectories for the state variables ($X(t),\,R(t)$, middle) and the full drop profile ($h(x,t)$, right), obtained by numerical optimization for small surface tension or large active capillary number ($\gamma=0.15$, Ca$_\zeta=383.66$). The drop adopts a strongly asymmetric shape, with an advancing peak and receding tail, like in Ref.~\cite{loisy2020many}. (B) Similar plots with the optimal activity controls ($\zeta_0(t),\,\Delta\zeta(t)$, left), and corresponding drop trajectory ($X(t),\,R(t)$, middle; $h(x,t)$, right), now obtained for large surface tension or small active capillary number ($\gamma=2$, Ca$_\zeta=30.91$). The transport plan fares poorly as the drops fails to reach the desired final position and size, and wastes a significant amount of energy in futile size oscillations ($R(t)$, middle) that don't aid in transport. Both (A-B) are computed using $X_T=0.8$ and $R_T=3$ ($X_0=0$ and $R_0=\sqrt{6}$ is kept fixed throughout), though similar policies are obtained for other values of $X_T/R_T$ as well. (C-E) The total cost ($\mathcal{C}_{\rm opt}$, C), efficiency [$\e_W$ (Eqs.~\ref{eq:Wmin},~\ref{eq:eff}), D], and precision ($X(T)/R(T)$, E) of the numerically computed optimal transport protocol plotted against Ca$_\zeta$, for different tasks labeled by increasing $X_T/R_T$ (blue to green). Remarkably, the performance curves present an optimal trade-off in balancing active forces against passive ones to attain improved drop transport plans at intermediate values of Ca$_\zeta$.}
	\label{fig:pde}
\end{figure*}
As the PDE model permits complex shape change of the drop, we lift the restriction of the controlled dynamics on the slow manifold selected by capillarity and allow aritrary shape variations by tuning $\gamma$ (conversely Ca$_\zeta$, Eq.~\ref{eq:Ca}). In order to select smooth control policies, we generalize the cost function $\mathcal{C}=\mathcal{W}+\mathcal{T}+\mathcal{R}$ (SI) to include a regularizing term $\mathcal{R}$ in addition to the total dissipation ($\mathcal{W}$, Eq.~\ref{eq:W}) and a finite terminal cost ($\mathcal{T}$ with $\mu_X,\mu_R<\infty$, Eq.~\ref{eq:T}). The regularization cost penalizes temporal jumps or changes in activity (see SI) and it implements a discretized version of \emph{minimal attention control} \cite{brockett1997minimum}.
In the presence of a finite surface tension $\gamma$, we must also correctly implement contact line motion at the boundary of the drop, which we do by simply introducing a thin precursor film that coats the entire surface and a disjoining potential \cite{de1985wetting} that sets both the film thickness and the equilibrium contact angle of a sessile drop (see SI for details). These additional terms are negligible in the bulk of the drop, but are dominant near the contact lines, which will become important when active and capillary forces are comparable (Ca$_\zeta\lesssim\mathcal{O}(10)$), similar to recently explored scenarios the context of steady migration of cells \cite{loisy2020many} in the absence of any dynamic control.

We numerically integrate the dynamical equations using a finite element method implemented using the FEniCS open source package \cite{alnaes2015fenics,logg2012automated} and perform constrained optimization using a gradient-free covariance matrix adaptation evolution strategy (CMA-ES) \cite{hansen2006cma} (see SI for details). Multiple runs are sequentially minimized with independent initializations for the activity profile and we choose the lowest cost solution as an estimate for the optimum. As before, the drop is initially at the origin ($X_0=0$) with a size $R_0=\sqrt{6}$, which corresponds to an equilibrium contact angle of $\phi_{\rm eq}=\pi/4$. The viscosity $\eta=0.1$ and total time $T=1$ are fixed in all the runs. To probe the caliber of the optimal policies obtained across tasks of increasing difficulty, we vary the imposed drop translation $X_T$ along with the surface tension $\gamma$.

For small surface tension or strong active driving (large Ca$_\zeta$), the dynamics of the drop is dominated by bulk dissipation. A typical trajectory of the drop profile (see Movie S1) and the controls is shown in Fig.~\ref{fig:pde}A, for $X_T=0.8$ and Ca$_\zeta\sim383$. The drop develops a prominent advancing peak and a thin receding tail (Fig.~\ref{fig:pde}A, right and Movie S1), similar to steadily translating drop shapes recently obtained in Ref.~\cite{loisy2020many}. The optimal controls and the drop motion vary smoothly (Fig.~\ref{fig:pde}A, left and middle) and successfully accomplish the transport task. Although the drop undergoes dramatic shape changes, the optimal controls are qualitatively consistent with our ODE results (Fig.~\ref{fig:sym}A-B). In particular, we note that the mean activity ($\zeta_0$) changes sign, switching from contractile to extensile activity, as predicted by our reduced order description (Fig.~\ref{fig:sym}A). Larger values of surface tension (smaller Ca$_\zeta$) lead to a qualitatively different scenario. As shown in Fig.~\ref{fig:pde}B and Movie S2, for $X_T=0.8$ and Ca$_\zeta\sim 31$, the drop remains nearly stationary for a finite time period, after which it advances by a small amount. This is reminiscent of ``waiting-time'' solutions \cite{lacey1982waiting} that are present in nonlinear diffusion equations of the form we have here. The drop performs rapid small scale oscillations of its shape and size (Fig.~\ref{fig:pde}B, middle and right; Movie S2) that arise from a competition between the active and passive (capillary) forces. While the active controls drive droplet motion, surface tension and substrate wetting resist variations in the drop shape and the contact angle, thereby limiting the translation achieved by the drop. The resulting ``futile'' oscillations wastefully dissipate large amounts of energy, performing poorly in the transport task (Fig.~\ref{fig:pde}B, left and middle).

The varied optimal strategies obtained upon tuning surface tension suggests a potential trade-off between active and passive forces. While large Ca$_\zeta$ allows for robust and efficient transport, it also generates dramatic shape changes which dissipate excessive energy. Smaller Ca$_\zeta$ restricts shape change, but consequently dissipates energy into futile oscillations that fail to complete the transport task. This suggests that an intermediate surface tension or Ca$_\zeta$ would serve as the best choice to tune the optimal transport plan by balancing active and passive forces. To quantify this trade-off, we employ three different performance metrics as a function of Ca$_\zeta$ and the task difficulty characterized by increasing $X_T/R_T$. The first is simply the total cost of the optimal solution ($\mathcal{C}_{\rm opt}$), plotted in Fig.~\ref{fig:pde}C. Second, we use a simple measure of efficiency ($\e_W$) to quantify the excess dissipation in the optimal solution (Fig.~\ref{fig:pde}D). We note that, for any arbitrary drop trajectory starting at the origin, there is a minimal amount of work that must \emph{necessarily} be expended, given by (see SI for a derivation)
\begin{equation}
	\mathcal{W}_{\rm min}=\dfrac{3\eta}{2T\lVert h\rVert_{\infty}^2}\left[X(T)^2+\dfrac{(\Delta(T)-\Delta(0))^2}{4\langle\Delta\rangle_T}\right]\;,
	\label{eq:Wmin}
\end{equation}
where $\lVert h\rVert_\infty=\sup_{x,t}h(x,t)$ is the maximum value of the drop height attained throughout its trajectory and $\Delta(t)=\int\dd x\;(x-X(t))^2h(x,t)$ is the variance in the drop height, which is related to the size of the drop ($\Delta\propto R^2$). Note that Eq.~\ref{eq:Wmin} is independent of the rheology and constitutive law for the fluid stress, and simply relies on the existence of a flux $q(x,t)$ that directly determines the dissipation (see SI). As the work done in the optimal solution is bounded below by construction ($\mathcal{W}_{\rm opt}\geq \mathcal{W}_{\rm min}$), we define the mechanical efficiency for the optimal solution
\begin{equation}
	\e_{W}=\dfrac{\mathcal{W}_{\rm min}}{\mathcal{W}_{\rm opt}}\leq 1\;,\label{eq:eff}
\end{equation}
which quantifies the extent to which energy is dissipated in internal modes rather than transporting the drop (Fig.~\ref{fig:pde}D). Finally, we also use the coefficient of variation of the height, i.e., the ratio of the final displacement to size achieved by the drop, $X(T)/R(T)$, as a measure of transport precision (Fig.~\ref{fig:pde}E, solid), and compare it against the prescribed value $X_T/R_T$ for a given transport task (Fig.~\ref{fig:pde}E, dashed). Interestingly, the optimal solutions for $X_T/R_T=0.2,\,0.27$ achieve a higher precision (solid curves, Fig.~\ref{fig:pde}E) than required by the task (dashed lines, Fig.~\ref{fig:pde}E) across a large range of Ca$_\zeta$, but this enhanced performance degrades for larger $X_T/R_T=0.33,\,0.4$. As anticipated, all three performance metrics are nonmonotonic and display an optimal trade-off at intermediate values of Ca$_\zeta$ (Fig.~\ref{fig:pde}C-E), with both the efficiency and the precision being maximized around Ca$_\zeta\sim 100-200$, while the optimal cost is minimal at a slightly higher Ca$_\zeta\sim250-380$. Qualitatively similar trends appear as we vary the transport task via $X_T/R_T$ (Fig.~\ref{fig:pde}C-E), though for larger values of $X_T/R_T=0.33,\,0.4$ (in the infeasible region of the symmetric ODE optimal transport, Fig.~\ref{fig:sym}C), the performance curves exhibit more kinks, perhaps suggestive of a rougher cost landscape with many nearly degenerate local minima when the task difficulty increases.

\section*{Discussion}
Complementing classical optimal transportation and its hydrodynamic analogies that use a very specific form of the cost \cite{benamou2000computational}, we have formulated a new framework to address questions of optimal mass transport in physical continua obeying complex dynamical constraints, and illustrated its utility in a simple yet rich problem of transporting a drop of an active suspension by dissipating the least amount of energy. Our strategy combines a finite dimensional (ODE) description based on a physically motivated mode reduction scheme, along with the full infinite dimensional (PDE) model, both of which we couch within optimal control theory to obtain a tractable and interpretable characterization of the resulting optimal transport policies. An important outcome is the prediction of intuitive ``gather-move-spread'' style strategies and simple trade-offs between active and passive forces that emerge naturally within our formulation of optimal drop transport, with implications for a wide range of synthetic and living active matter.

In synthetic systems, digital microfluidics \cite{fair2007digital} and bicomponent volatile liquids \cite{cira2015vapour} might provide an immediate platform to deploy our framework, while in a the biological context, our work is directly relevant to the control of self-propelled drops composed of microtubule-kinesin nematic gels \cite{sanchez2012spontaneous}, swimming bacteria \cite{rajabi2020directional}, and isolated motile cells \cite{keren2008mechanism,bruckner2019stochastic} that are often viewed as active drops \cite{kruse2006contractility,recho2013contraction,tjhung2015minimal,loisy2019tractionless,*loisy2020many}. Our results could also be tested using optogenetically controlled living motile cells \cite{wu2009genetically} or in reconstituted active drops \cite{boukellal2004soft,sanchez2012spontaneous}. Since the cortical tension of individual cells varies in the range of $\gamma\sim0.1-1~$mN/m \cite{salbreux2012actin}, and the characteristic active stress $\zeta h\sim 1~$kPa \cite{kruse2006contractility}, Ca$_\zeta\sim 10^2-10^3$ for a $R=10~\mu$m size cell (assuming an average height $h\sim1~\mu$m), allowing for an exploration of the transport cost versus internal dissipation trade-off at intermediate Ca$_\zeta\sim\mathcal{O}(100)$ that we have uncovered here. This suggests that contractility driven cellular motility may be optimal even beyond steady translation \cite{recho2014optimality} by harnessing dynamic optimal protocols and internally regulating differential contractility against surface tension. Extensions of our framework can also easily be used to address the control and patterning of localized structures such as defects in active fluids, which has been the focus of much research in recent years \cite{shankar2019hydrodynamics,ross2019controlling,zhang2021spatiotemporal,norton2020optimal}. 

More broadly, beyond the control of active systems, our formulation of optimal transport offers an alternative choice of transport metrics that are physically motivated and potentially richer than the conventional $L^2$-Wasserstein norm common to standard optimal transport \cite{villani2008optimal}, yet their mathematical properties remain to be uncovered. A tantalizing possibility is to exploit thermodynamic analogies connecting minimum dissipation protocols in stochastic systems to optimal transport \cite{aurell2011optimal}. In this regard, a generalization of our framework to include fluctuations within stochastic optimal control \cite{chen2020optimal} would be a promising future direction.
%


\acknow{
We acknowledge partial financial support from  the Harvard Society of Fellows (SS), the Harvard MRSEC DMR-2011754 (LM), The NSF Simons Center for Mathematics and Complex Biological Systems DMR-1764269 (LM) the Simons Foundation (LM) and the Seydoux Fund (LM) and gratefully acknowledges illuminating discussions during the virtual 2020 KITP program on ``Symmetry, Thermodynamics and Topology in Active Matter'', which was supported in part by the National Science Foundation under Grant No.~NSF PHY-1748958.
}

\showacknow{} 


\begin{thebibliography}{10}

\bibitem{needleman2017active}
D Needleman, Z Dogic, Active matter at the interface between materials science
  and cell biology.
\newblock {\em\protect\JournalTitle{Nature Reviews Materials}} \textbf{2},
  1--14 (2017).

\bibitem{zhang2021autonomous}
R Zhang, A Mozaffari, JJ de~Pablo, Autonomous materials systems from active
  liquid crystals.
\newblock {\em\protect\JournalTitle{Nature Reviews Materials}} \textbf{6},
  437--453 (2021).

\bibitem{shankar2020topological}
S Shankar, A Souslov, MJ Bowick, MC Marchetti, V Vitelli, Topological active
  matter (2020).

\bibitem{sanchez2012spontaneous}
T Sanchez, DT Chen, SJ DeCamp, M Heymann, Z Dogic, Spontaneous motion in
  hierarchically assembled active matter.
\newblock {\em\protect\JournalTitle{Nature}} \textbf{491}, 431--434 (2012).

\bibitem{bricard2013emergence}
A Bricard, JB Caussin, N Desreumaux, O Dauchot, D Bartolo, Emergence of
  macroscopic directed motion in populations of motile colloids.
\newblock {\em\protect\JournalTitle{Nature}} \textbf{503}, 95--98 (2013).

\bibitem{zhou2014living}
S Zhou, A Sokolov, OD Lavrentovich, IS Aranson, Living liquid crystals.
\newblock {\em\protect\JournalTitle{Proceedings of the National Academy of
  Sciences}} \textbf{111}, 1265--1270 (2014).

\bibitem{duclos2018spontaneous}
G Duclos, et~al., Spontaneous shear flow in confined cellular nematics.
\newblock {\em\protect\JournalTitle{Nature physics}} \textbf{14}, 728--732
  (2018).

\bibitem{marchetti2013hydrodynamics}
MC Marchetti, et~al., Hydrodynamics of soft active matter.
\newblock {\em\protect\JournalTitle{Reviews of Modern Physics}} \textbf{85},
  1143 (2013).

\bibitem{vizsnyiczai2017light}
G Vizsnyiczai, et~al., Light controlled 3d micromotors powered by bacteria.
\newblock {\em\protect\JournalTitle{Nature communications}} \textbf{8}, 1--7
  (2017).

\bibitem{krishnamurthy2016micrometre}
S Krishnamurthy, S Ghosh, D Chatterji, R Ganapathy, A Sood, A micrometre-sized
  heat engine operating between bacterial reservoirs.
\newblock {\em\protect\JournalTitle{Nature Physics}} \textbf{12}, 1134--1138
  (2016).

\bibitem{arlt2018painting}
J Arlt, VA Martinez, A Dawson, T Pilizota, WC Poon, Painting with light-powered
  bacteria.
\newblock {\em\protect\JournalTitle{Nature communications}} \textbf{9}, 1--7
  (2018).

\bibitem{frangipane2018dynamic}
G Frangipane, et~al., Dynamic density shaping of photokinetic e. coli.
\newblock {\em\protect\JournalTitle{Elife}} \textbf{7}, e36608 (2018).

\bibitem{ross2019controlling}
TD Ross, et~al., Controlling organization and forces in active matter through
  optically defined boundaries.
\newblock {\em\protect\JournalTitle{Nature}} \textbf{572}, 224--229 (2019).

\bibitem{zhang2021spatiotemporal}
R Zhang, et~al., Spatiotemporal control of liquid crystal structure and
  dynamics through activity patterning.
\newblock {\em\protect\JournalTitle{Nature Materials}} \textbf{20}, 875--882
  (2021).

\bibitem{turiv2020topology}
T Turiv, et~al., Topology control of human fibroblast cells monolayer by liquid
  crystal elastomer.
\newblock {\em\protect\JournalTitle{Science Advances}} \textbf{6}, eaaz6485
  (2020).

\bibitem{endresen2019topological}
KD Endresen, M Kim, M Pittman, Y Chen, F Serra, Topological defects of integer
  charge in cell monolayers.
\newblock {\em\protect\JournalTitle{Soft Matter}} \textbf{17}, 5878--5887
  (2021).

\bibitem{cohen2014galvanotactic}
DJ Cohen, WJ Nelson, MM Maharbiz, Galvanotactic control of collective cell
  migration in epithelial monolayers.
\newblock {\em\protect\JournalTitle{Nature materials}} \textbf{13}, 409--417
  (2014).

\bibitem{zajdel2020scheepdog}
TJ Zajdel, G Shim, L Wang, A Rossello-Martinez, DJ Cohen, Scheepdog:
  programming electric cues to dynamically herd large-scale cell migration.
\newblock {\em\protect\JournalTitle{Cell systems}} \textbf{10}, 506--514
  (2020).

\bibitem{monge1781memoire}
G Monge, M{\'e}moire sur la th{\'e}orie des d{\'e}blais et des remblais.
\newblock {\em\protect\JournalTitle{Histoire de l'Acad{\'e}mie Royale des
  Sciences de Paris}}, 666--704 (1781).

\bibitem{kantorovich1942translocation}
LV Kantorovich, On the translocation of masses in {\em Dokl.~Akad.~ Nauk.~USSR
  (NS)}.
\newblock Vol.{}~37, pp. 199--201 (1942) English translation: J.~Math.~Sci.,
  133, 4 (2006), 1381-1382.

\bibitem{villani2008optimal}
C Villani, {\em Optimal transport: old and new}.
\newblock (Springer Science \& Business Media) Vol.{} 338, (2008).

\bibitem{benamou2000computational}
JD Benamou, Y Brenier, A computational fluid mechanics solution to the
  monge-kantorovich mass transfer problem.
\newblock {\em\protect\JournalTitle{Numerische Mathematik}} \textbf{84},
  375--393 (2000).

\bibitem{joanny_ramaswamy_2012}
JF Joanny, S Ramaswamy, A drop of active matter.
\newblock {\em\protect\JournalTitle{Journal of Fluid Mechanics}} \textbf{705},
  46–57 (2012).

\bibitem{loisy2019tractionless}
A Loisy, J Eggers, TB Liverpool, Tractionless self-propulsion of active drops.
\newblock {\em\protect\JournalTitle{Physical review letters}} \textbf{123},
  248006 (2019).

\bibitem{loisy2020many}
A Loisy, J Eggers, TB Liverpool, How many ways a cell can move: the modes of
  self-propulsion of an active drop.
\newblock {\em\protect\JournalTitle{Soft matter}} \textbf{16}, 3106--3124
  (2020).

\bibitem{simha2002hydrodynamic}
RA Simha, S Ramaswamy, Hydrodynamic fluctuations and instabilities in ordered
  suspensions of self-propelled particles.
\newblock {\em\protect\JournalTitle{Physical review letters}} \textbf{89},
  058101 (2002).

\bibitem{de1985wetting}
PG De~Gennes, Wetting: statics and dynamics.
\newblock {\em\protect\JournalTitle{Reviews of modern physics}} \textbf{57},
  827 (1985).

\bibitem{de2013capillarity}
PG De~Gennes, F Brochard-Wyart, D Qu{\'e}r{\'e}, {\em Capillarity and wetting
  phenomena: drops, bubbles, pearls, waves}.
\newblock (Springer Science \& Business Media), (2013).

\bibitem{pietzonka2019autonomous}
P Pietzonka, {\'E} Fodor, C Lohrmann, ME Cates, U Seifert, Autonomous engines
  driven by active matter: Energetics and design principles.
\newblock {\em\protect\JournalTitle{Physical Review X}} \textbf{9}, 041032
  (2019).

\bibitem{recho2014optimality}
P Recho, JF Joanny, L Truskinovsky, Optimality of contraction-driven crawling.
\newblock {\em\protect\JournalTitle{Physical Review Letters}} \textbf{112},
  218101 (2014).

\bibitem{liberzon2011calculus}
D Liberzon, {\em Calculus of variations and optimal control theory: a concise
  introduction}.
\newblock (Princeton university press), (2011).

\bibitem{hansen2006cma}
N Hansen, {\em The CMA Evolution Strategy: A Comparing Review}, eds.{} JA
  Lozano, P Larra{\~{n}}aga, I Inza, E Bengoetxea.
\newblock (Springer Berlin Heidelberg, Berlin, Heidelberg), pp. 75--102 (2006).

\bibitem{recho2013contraction}
P Recho, T Putelat, L Truskinovsky, Contraction-driven cell motility.
\newblock {\em\protect\JournalTitle{Physical review letters}} \textbf{111},
  108102 (2013).

\bibitem{brockett1976nonlinear}
RW Brockett, Nonlinear systems and differential geometry.
\newblock {\em\protect\JournalTitle{Proceedings of the IEEE}} \textbf{64},
  61--72 (1976).

\bibitem{hermann1977nonlinear}
R Hermann, A Krener, Nonlinear controllability and observability.
\newblock {\em\protect\JournalTitle{IEEE Transactions on automatic control}}
  \textbf{22}, 728--740 (1977).

\bibitem{pontryagin2018mathematical}
L Pontryagin, {\em Mathematical Theory of Optimal Processes}.
\newblock (CRC Press), (2018).

\bibitem{andersson2019casadi}
JA Andersson, J Gillis, G Horn, JB Rawlings, M Diehl, Casadi: a software
  framework for nonlinear optimization and optimal control.
\newblock {\em\protect\JournalTitle{Mathematical Programming Computation}}
  \textbf{11}, 1--36 (2019).

\bibitem{brockett1997minimum}
W Brockett, Minimum attention control in {\em Proceedings of the 36th IEEE
  Conference on Decision and Control}.
\newblock (IEEE), Vol.{}~3, pp. 2628--2632 (1997).

\bibitem{alnaes2015fenics}
M Aln{\ae}s, et~al., The fenics project version 1.5.
\newblock {\em\protect\JournalTitle{Archive of Numerical Software}} \textbf{3}
  (2015).

\bibitem{logg2012automated}
A Logg, KA Mardal, G Wells, {\em Automated solution of differential equations
  by the finite element method: The FEniCS book}.
\newblock (Springer Science \& Business Media) Vol.{}~84, (2012).

\bibitem{lacey1982waiting}
A Lacey, J Ockendon, A Tayler, “waiting-time” solutions of a nonlinear
  diffusion equation.
\newblock {\em\protect\JournalTitle{SIAM Journal on Applied Mathematics}}
  \textbf{42}, 1252--1264 (1982).

\bibitem{fair2007digital}
RB Fair, Digital microfluidics: is a true lab-on-a-chip possible?
\newblock {\em\protect\JournalTitle{Microfluidics and Nanofluidics}}
  \textbf{3}, 245--281 (2007).

\bibitem{cira2015vapour}
NJ Cira, A Benusiglio, M Prakash, Vapour-mediated sensing and motility in
  two-component droplets.
\newblock {\em\protect\JournalTitle{Nature}} \textbf{519}, 446--450 (2015).

\bibitem{rajabi2020directional}
M Rajabi, H Baza, T Turiv, OD Lavrentovich, Directional self-locomotion of
  active droplets enabled by nematic environment.
\newblock {\em\protect\JournalTitle{Nature Physics}} \textbf{17}, 260--266
  (2021).

\bibitem{keren2008mechanism}
K Keren, et~al., Mechanism of shape determination in motile cells.
\newblock {\em\protect\JournalTitle{Nature}} \textbf{453}, 475--480 (2008).

\bibitem{bruckner2019stochastic}
DB Br{\"u}ckner, et~al., Stochastic nonlinear dynamics of confined cell
  migration in two-state systems.
\newblock {\em\protect\JournalTitle{Nature Physics}} \textbf{15}, 595--601
  (2019).

\bibitem{kruse2006contractility}
K Kruse, J Joanny, F J{\"u}licher, J Prost, Contractility and retrograde flow
  in lamellipodium motion.
\newblock {\em\protect\JournalTitle{Physical biology}} \textbf{3}, 130 (2006).

\bibitem{tjhung2015minimal}
E Tjhung, A Tiribocchi, D Marenduzzo, M Cates, A minimal physical model
  captures the shapes of crawling cells.
\newblock {\em\protect\JournalTitle{Nature communications}} \textbf{6}, 1--9
  (2015).

\bibitem{wu2009genetically}
YI Wu, et~al., A genetically encoded photoactivatable rac controls the motility
  of living cells.
\newblock {\em\protect\JournalTitle{Nature}} \textbf{461}, 104--108 (2009).

\bibitem{boukellal2004soft}
H Boukellal, O Camp{\'a}s, JF Joanny, J Prost, C Sykes, Soft listeria:
  actin-based propulsion of liquid drops.
\newblock {\em\protect\JournalTitle{Physical Review E}} \textbf{69}, 061906
  (2004).

\bibitem{salbreux2012actin}
G Salbreux, G Charras, E Paluch, Actin cortex mechanics and cellular
  morphogenesis.
\newblock {\em\protect\JournalTitle{Trends in cell biology}} \textbf{22},
  536--545 (2012).

\bibitem{shankar2019hydrodynamics}
S Shankar, MC Marchetti, Hydrodynamics of active defects: from order to chaos
  to defect ordering.
\newblock {\em\protect\JournalTitle{Physical Review X}} \textbf{9}, 041047
  (2019).

\bibitem{norton2020optimal}
MM Norton, P Grover, MF Hagan, S Fraden, Optimal control of active nematics.
\newblock {\em\protect\JournalTitle{Physical review letters}} \textbf{125},
  178005 (2020).

\bibitem{aurell2011optimal}
E Aurell, C Mej{\'\i}a-Monasterio, P Muratore-Ginanneschi, Optimal protocols
  and optimal transport in stochastic thermodynamics.
\newblock {\em\protect\JournalTitle{Physical review letters}} \textbf{106},
  250601 (2011).

\bibitem{chen2020optimal}
Y Chen, TT Georgiou, M Pavon, Optimal transport in systems and control.
\newblock {\em\protect\JournalTitle{Annual Review of Control, Robotics, and
  Autonomous Systems}} \textbf{4} (2020).

\end{thebibliography}

\end{document}